\documentclass[sigconf]{acmart}

\AtBeginDocument{%
  \providecommand\BibTeX{{%
    \normalfont B\kern-0.5em{\scshape i\kern-0.25em b}\kern-0.8em\TeX}}}

\setcopyright{acmcopyright}
\acmConference{DLP-KDD 2021}{August 15, 2021}{Singapore}

\begin{document}

\title{MaskNet: Introducing Feature-Wise Multiplication to CTR Ranking Models by Instance-Guided Mask}

\author{Zhiqiang Wang, Qingyun She, Junlin Zhang}
\affiliation{
     \institution{Sina Weibo Corp}
     \city{Beijing}
     \country{China}}
\email{roky2813@sina.com,qingyun_she@163.com,junlin6@staff.weibo.com}


\begin{abstract}
Click-Through Rate(CTR) estimation has become one of the most fundamental tasks in many real-world applications and it’s important for ranking models to effectively capture complex high-order features. Shallow feed-forward network is widely used in many state-of-the-art DNN models such as FNN, DeepFM and xDeepFM to implicitly capture high-order feature interactions. However, some research has proved that  addictive feature interaction, particular feed-forward neural networks, is inefficient in capturing common feature interaction. To resolve this problem, we introduce specific multiplicative operation into DNN ranking system by proposing instance-guided mask which performs element-wise product both on the feature embedding and feed-forward layers guided by input instance. We also turn the feed-forward layer in DNN model into a mixture of addictive and multiplicative feature interactions by proposing MaskBlock in this paper. MaskBlock combines the layer normalization, instance-guided mask, and feed-forward layer and it is a basic building block to be used to design new ranking model under various configurations. The model consisting of MaskBlock is called MaskNet in this paper and two new MaskNet models are proposed to show the effectiveness of MaskBlock as basic building block for composing high performance ranking systems.  The experiment results on three real-world datasets demonstrate that our proposed MaskNet models outperform state-of-the-art models such as DeepFM and xDeepFM significantly, which implies MaskBlock is an effective basic building unit for composing new high performance ranking systems.

\end{abstract}



\maketitle

\section{Introduction}
Click-through rate (CTR) prediction is to predict the probability of a user clicking on the recommended items. It plays important role in personalized advertising and recommender systems. Many models have been proposed to resolve this problem such as Logistic Regression (LR) \cite{10.1145/2487575.2488200}, Polynomial-2 (Poly2) \cite{rendle2010factorization}, tree-based models \cite{he2014practical}, tensor-based models \cite{koren2009matrix}, Bayesian models \cite{graepel2010web}, and Field-aware Factorization Machines (FFMs) \cite{juan2016field}. In recent years, employing DNNs for CTR estimation has also been a research trend in this field and some deep learning based models have been introduced such as Factorization-Machine Supported Neural Networks(FNN)\cite{zhang2016deep}, Attentional  Factorization Machine (AFM)\cite{cheng2016wide}, wide \& deep(W\&D)\cite{xiao2017attentional}, DeepFM\cite{guo2017deepfm}, xDeepFM\cite{lian2018xdeepfm} etc.

Feature interaction is critical for CTR tasks and it's important for ranking model to effectively capture these complex features. Most DNN ranking models such as FNN , W\&D, DeepFM and xDeepFM use the shallow MLP layers to model high-order interactions in an implicit way and it's an important component in current state-of-the-art ranking systems.

However, Alex Beutel et.al \cite{beutel2018latent} have proved that addictive feature interaction, particular feed-forward neural networks, is inefficient in capturing common feature crosses. They proposed a simple but effective approach named "latent cross" which is a kind of multiplicative interactions between the context embedding and the neural network hidden states in RNN model. Recently, Rendle et.al's work \cite{rendle2020neural} also shows that a carefully configured dot product baseline largely outperforms the MLP layer in collaborative filtering. While a MLP can in theory approximate any function, they show that it is non-trivial to learn a dot product with an MLP and  learning a dot product with high accuracy for a decently large embedding dimension requires a large model capacity as well as many training data. Their work also proves the inefficiency of MLP layer's ability to model complex feature interactions.

Inspired by "latent cross"\cite{beutel2018latent} and Rendle's work \cite{rendle2020neural},  we care about the following question: Can we improve the DNN ranking systems by introducing specific multiplicative operation into it to make it efficiently capture complex feature interactions?

In order to overcome the problem of inefficiency of feed-forward layer to capture complex feature cross, we introduce a special kind of multiplicative operation into DNN ranking system in this paper. First, we propose an instance-guided mask performing element-wise production on the feature embedding and feed-forward layer. The instance-guided mask utilizes the global information collected from input instance to dynamically highlight the informative elements in feature embedding and hidden layer in a unified manner. There are two main advantages for adopting the instance-guided mask: firstly, the element-wise product between the mask and hidden layer or feature embedding layer brings multiplicative operation into DNN ranking system in unified way to more efficiently capture complex feature interaction. Secondly, it’s a kind of fine-gained bit-wise attention guided by input instance which can both weaken the influence of noise in feature embedding and MLP layers while highlight the informative signals in DNN ranking systems.

By combining instance-guided mask, a following feed-forward layer and layer normalization, MaskBlock is proposed by us to turn the commonly used feed-forward layer into a mixture of addictive and multiplicative feature interactions. The instance-guided mask introduces multiplicative interactions and the following feed-forward hidden layer aggregates the masked  information in order to better capture the important feature interactions. While the layer normalization can ease optimization of the network.

MaskBlock can be regarded as a  basic building block to design new ranking models under some kinds of configuration. The model consisting of MaskBlock is called MaskNet in this paper and two new MaskNet models are proposed to show the effectiveness of MaskBlock as basic building block for composing high performance ranking systems.

The contributions of our work are summarized as follows:
\begin{enumerate}
    \item In this work, we  propose an instance-guided mask performing element-wise product both on the feature embedding and feed-forward layers in DNN models. The global context information contained in the instance-guided mask is dynamically incorporated into the feature embedding and feed-forward layer to highlight the important elements.

    \item We propose a  basic building block named MaskBlock which  consists of three key components: instance-guided mask, a  following feed-forward hidden layer and layer normalization module. In this way, we turn the widely used feed-forward layer of a standard DNN model into a mixture of addictive and multiplicative feature interactions.

    \item We also propose a new ranking framework named  MaskNet to compose new ranking system by utilizing MaskBlock as basic building unit. To be more specific, the serial MaskNet model and parallel MaskNet model are designed based on the MaskBlock in this paper. The serial rank model stacks MaskBlock block by block while the parallel rank model puts many MaskBlocks in parallel on a sharing feature embedding layer.

    \item Extensive experiments are conduct on three real-world datasets and the experiment results demonstrate that our proposed two MaskNet models outperform state-of-the-art models significantly. The results imply MaskBlock indeed  enhance DNN model’s  ability of capturing complex feature interactions through introducing multiplicative operation into DNN models by instance-guided mask.

\end{enumerate}

The rest of this paper is organized as follows. Section 2 introduces some related works which are relevant with our proposed model. We introduce our proposed models in detail in Section 3. The experimental results on three real world datasets are presented and discussed in Section 4. Section 5 concludes our work in this paper.

\section{Related Work}
\subsection{State-Of-The-Art CTR Models}

Many deep learning based CTR models have  been proposed in recent years and it is the key factor for most of these neural network based models to effectively model the feature interactions.

Factorization-Machine Supported Neural Networks (FNN)\cite{zhang2016deep} is a feed-forward neural network using FM to pre-train the embedding layer. Wide \& Deep Learning\cite{xiao2017attentional} jointly trains wide linear models and deep neural networks to combine the benefits of memorization and generalization for recommender systems. However, expertise feature engineering is still needed on the input to the wide part of Wide \& Deep model. To alleviate manual efforts in feature engineering, DeepFM\cite{guo2017deepfm} replaces the wide part of Wide \& Deep model with FM and shares the feature embedding between the FM and deep component.

While most DNN ranking models process high-order feature interactions by MLP layers in implicit way, some works explicitly introduce high-order feature interactions by sub-network. Deep \& Cross Network (DCN)\cite{wang2017deep} efficiently captures feature interactions of bounded degrees in an explicit fashion. Similarly, eXtreme Deep Factorization Machine (xDeepFM) \cite{lian2018xdeepfm} also models the low-order and high-order feature interactions in an explicit way by proposing a novel Compressed Interaction Network (CIN) part. AutoInt\cite{song2019autoint} uses a multi-head self-attentive neural network to explicitly model the feature interactions in the low-dimensional space. FiBiNET\cite{huang2019fibinet} can dynamically learn feature importance via the Squeeze-Excitation network (SENET) mechanism and feature interactions via bilinear function.

\subsection{Feature-Wise Mask Or Gating}
Feature-wise mask or gating  has been explored widely in vision \cite{hu2018squeeze,srivastava2015highway}, natural language processing \cite{dauphin2017language} and recommendation system\cite{ma2018modeling,ma2019hierarchical}. For examples, Highway Networks \cite{srivastava2015highway} utilize feature gating to ease gradient-based training of very deep networks. Squeeze-and-Excitation Networks\cite{hu2018squeeze} recalibrate feature responses by explicitly multiplying each channel with learned sigmoidal mask values. Dauphin et al.\cite{dauphin2017language} proposed gated linear unit (GLU) to utilize it to control what information should be propagated for predicting the next word in the language modeling task. Gating or mask mechanisms are also adopted in  recommendation systems. Ma et al. \cite{ma2018modeling} propose a novel multi-task learning approach, Multi-gate Mixture-of-Experts (MMoE), which explicitly learns to model task relationships from data. Ma et al.\cite{ma2019hierarchical} propose a hierarchical gating network (HGN) to capture both the long-term and short-term user interests. The feature gating and instance gating modules in HGN  select what item features can be passed to the downstream layers from the feature and instance levels, respectively.

\subsection{Normalization}
Normalization techniques have been recognized as very effective components in deep learning. Many normalization approaches have been proposed with the two most popular ones being BatchNorm \cite{ioffe2015batch} and LayerNorm \cite{ba2016layer} . Batch Normalization (Batch Norm or BN)\cite{ioffe2015batch} normalizes the features by the mean and variance computed within a mini-batch. Another example is layer normalization (Layer Norm or LN)\cite{ba2016layer} which was proposed to ease optimization of recurrent neural networks. Statistics of layer normalization are not computed across the $N$ samples in a mini-batch but are estimated in a layer-wise manner for each sample independently. Normalization methods have shown success in accelerating the training of deep networks.

\section{Our Proposed Model}


In this section, we first describe the feature embedding layer. Then the details of the instance-guided mask, MaskBlock and MaskNet structure we proposed will be introduced. Finally the log loss as a loss function is introduced.

\subsection{Embedding Layer}
The input data of CTR tasks usually consists of sparse and dense features and the sparse features are mostly categorical type. Such features are  encoded as one-hot vectors which often lead to excessively high-dimensional feature spaces for large vocabularies. The common solution to this problem is to introduce the embedding layer. Generally, the sparse input can be formulated as:
\begin{equation}
  x = [x_1, x_2, ..., x_f]
\end{equation}
where $f$ denotes the number of fields, and $x_i \in \mathbb{R}^n$ denotes a one-hot vector for a categorical  field with $n$ features and $x_i \in \mathbb{R}^n$ is vector with only one value for a numerical  field. We can obtain feature embedding $e_i$ for one-hot vector $x_i$ via:
\begin{equation}
  e_i = W_ex_i
\end{equation}
where $W_e \in \mathbb{R}^{k\times n}$ is the embedding matrix of $n$ features and $k$ is the dimension of field embedding. The numerical feature $x_j$ can also be converted into the same low-dimensional space by:
\begin{equation}
  e_j = V_jx_j
\end{equation}
where $V_j \in \mathbb{R}^k$ is the corresponding field embedding with size $k$.

Through the aforementioned method, an embedding layer is applied upon the raw feature input to compress it to a low dimensional, dense real-value vector. The result of embedding layer is a wide concatenated vector:

\begin{equation}
  \mathbf{V}_{emb} = concat(\mathbf{e}_1, \mathbf{e}_2, ..., \mathbf{e}_i, ..., \mathbf{e}_f)
\end{equation}
where $f$ denotes the number of fields, and  $\mathbf{e}_i \in \mathbb{R}^k$ denotes the embedding of one field. Although the feature lengths of input instances can be various, their embedding are of the same length $f \times k$, where $k$ is the dimension of field embedding.

We use instance-guided mask to introduce the multiplicative operation into DNN ranking system and here the so-called "instance" means the feature embedding layer of current input instance in the following part of this paper.

\subsection{Instance-Guided Mask}
We utilize the global information collected from input instance by instance-guided mask to dynamically highlight the informative elements in feature embedding and feed-forward layer. For feature embedding, the mask lays stress on the key elements with more information to effectively represent this feature. For the neurons in hidden layer, the mask helps those important feature interactions to stand out by considering the contextual information in the input instance. In addition to this advantage, the instance-guided mask also introduces the multiplicative operation into DNN ranking system to capture complex feature cross more efficiently.

\begin{figure}[!]
  \setlength{\abovecaptionskip}{1pt}
  \includegraphics[width=0.75\linewidth]{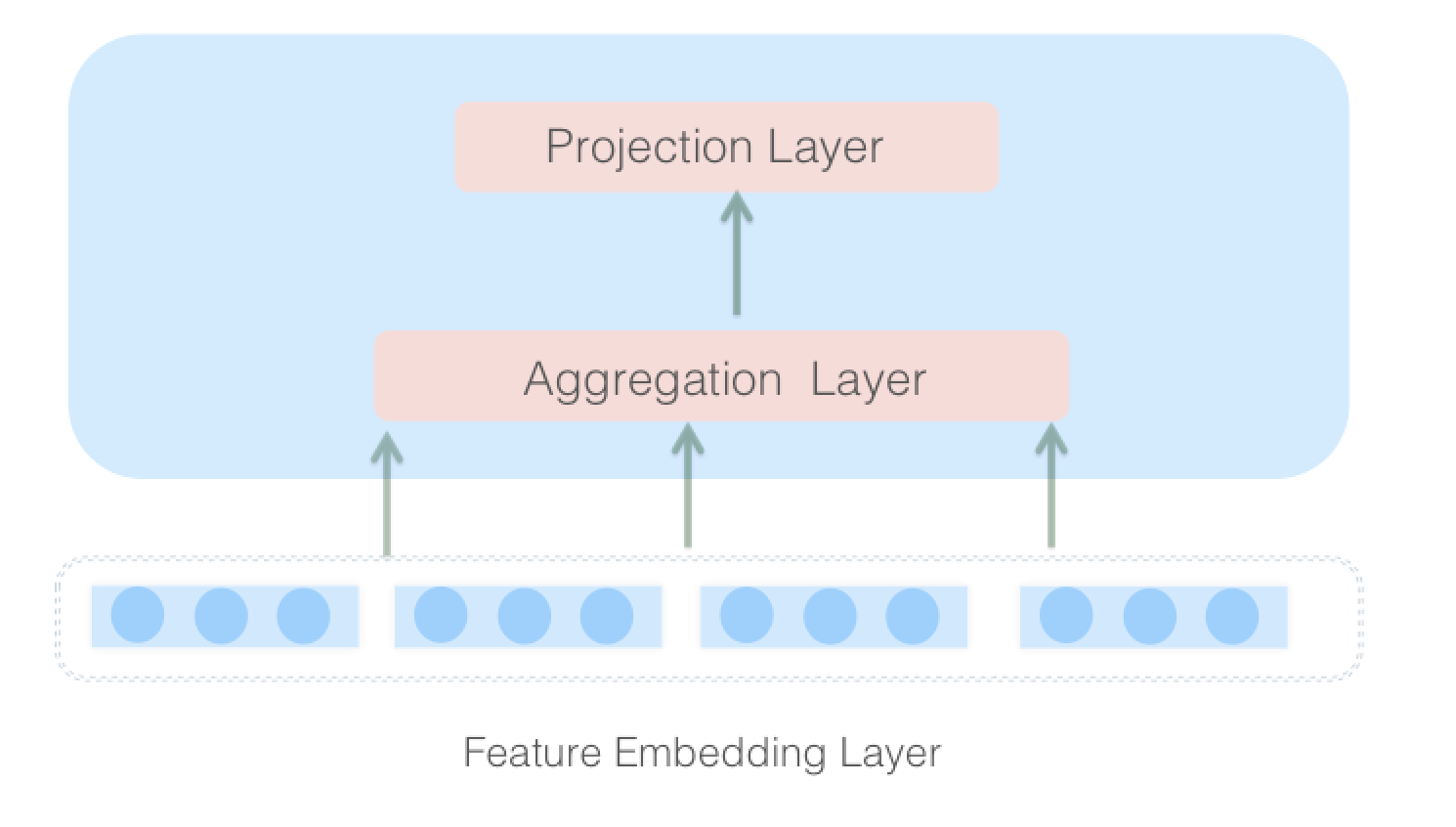}
  \caption{Neural Structure of Instance-Guided Mask}
  \label{Fig.Instance-Guided}
\end{figure}

 As depicted in Figure \ref{Fig.Instance-Guided}, two fully connected (FC) layers with identity function are used in instance-guided mask. Notice that the input of instance-guided mask is always from the input instance, that is to say, the feature embedding layer.

 The first FC layer is called "aggregation layer" and it is a relatively wider layer compared with the second FC layer in order to better collect the global  contextual information in input instance. The aggregation layer has parameters $W_{d1}$ and here $d$ denotes the $d$-th mask. For feature embedding and different MLP layers, we adopt different instance-guided mask owning its parameters to learn to capture various information for each layer from input instance.

The second FC layer named "projection layer" reduces dimensionality to the same size as feature embedding layer $V_{emb}$ or hidden layer $V_{hidden}$ with parameters $W_{d2}$, Formally,
\begin{equation}
  V_{mask} = W_{d2}(Relu(W_{d1}V_{emb} + \beta_{d1})) + \beta_{d2}
\end{equation}
where $V_{emb} \in \mathbb{R}^{m=f\times k}$ refers to the embedding layer of input instance, $W_{d1} \in \mathbb{R}^{t\times m}$ and $W_{d2} \in \mathbb{R}^{z\times t}$ are parameters for instance-guided mask, $t$ and $z$ respectively denotes the neural number of aggregation layer and projection layer, $f$ denotes the number of fields and $k$ is the dimension of field embedding. $\beta_{d1} \in \mathbb{R}^{t\times m}$ and $\beta_{d2} \in \mathbb{R}^{z\times t}$ are learned bias of the two FC layers. Notice here that the aggregation layer is usually wider than the projection layer because the size of the projection layer is required to be equal to the size of feature embedding layer or MLP layer. So we define the size $r=t/z$ as reduction ratio which is a hyper-parameter to control the ratio of neuron numbers of two layers.

Element-wise product is used in this work to incorporate the global contextual information aggregated by instance-guided mask into feature embedding or hidden layer as following:
\begin{equation}
  \begin{split}
  \mathbf{V}_{maskedEMB} &= \mathbf{V}_{mask} \odot \mathbf{V}_{emb} \\
  \mathbf{V}_{maskedHID} &= \mathbf{V}_{mask} \odot \mathbf{V}_{hidden}
\end{split}
\end{equation}
where $\mathbf{V}_{emb}$ denotes embedding layer and $\mathbf{V}_{hidden}$ means the feed-forward layer in DNN model, $\odot$ means the element-wise production between two vectors as follows:
\begin{equation}
  V_i \odot V_j = [V_{i1} \cdot V_{j1}, V_{i2} \cdot V_{j2}, ..., V_{iu} \cdot V_{ju}]
\end{equation}
here $u$ is the size of vector $V_i$ and $V_j$

The instance-guided mask can be regarded as a special kind of bit-wise attention or gating mechanism which uses the global context information contained in input instance to guide the parameter optimization during training. The bigger value in $V_{mask}$ implies that the model dynamically identifies an important element in feature embedding or hidden layer. It is used to boost the element in vector $V_{emb}$ or $V_{hidden}$. On the contrary, small value in $V_{mask}$ will suppress the uninformative elements or even noise by decreasing the values in the corresponding vector $V_{emb}$ or $V_{hidden}$.

The two main advantages in adopting the instance-guided mask are: firstly, the element-wise product between the mask and hidden layer or feature embedding layer brings multiplicative operation into DNN ranking system in unified way to more efficiently capture complex feature interaction. Secondly, this kind of fine-gained bit-wise attention guided by input instance can both weaken the influence of noise in feature embedding and MLP layers while highlight the informative signals in DNN ranking systems.

\subsection{MaskBlock}
\begin{figure}
  \setlength{\abovecaptionskip}{1pt}
  \includegraphics[width=0.65\linewidth]{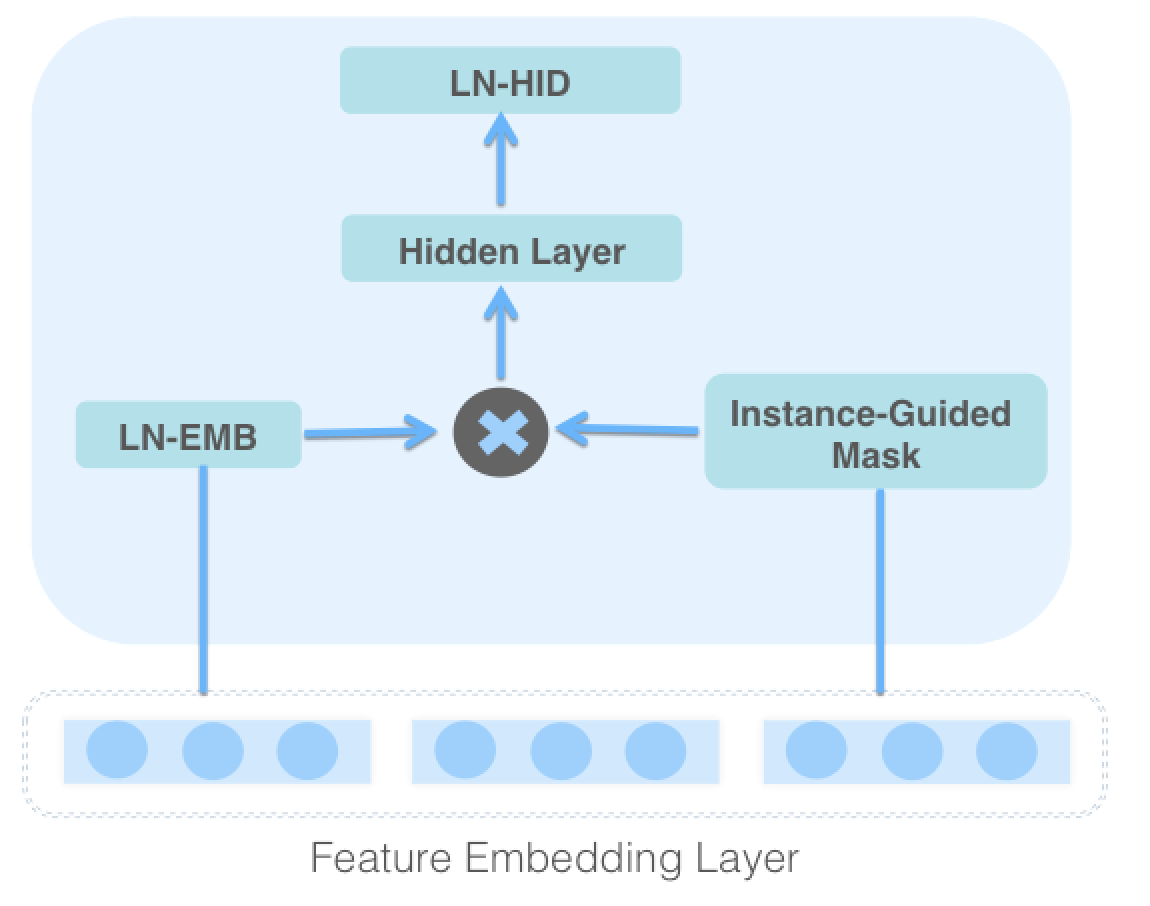}
  \caption{MaskBlock on Feature Embedding}
  \label{Fig.figure-2a}
\end{figure}

\begin{figure}
   \setlength{\abovecaptionskip}{1pt}
   \includegraphics[width=0.75\linewidth]{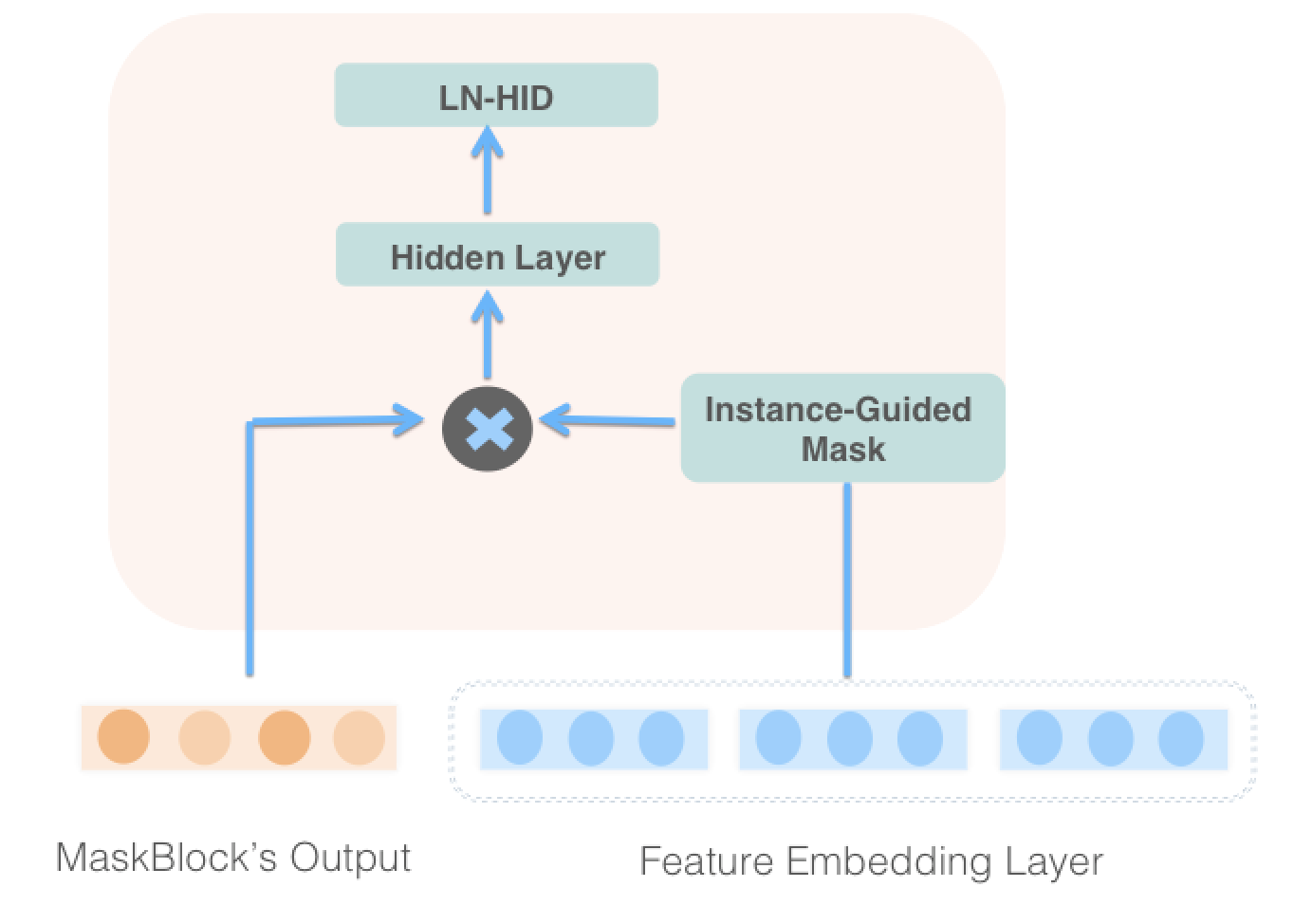}
   \caption{MaskBlock on MaskBlock }
   \label{Fig.figure-2b}

 \end{figure}

To overcome the problem of the inefficiency of feed-forward layer to capture complex feature interaction in DNN models, we propose a basic building block named MaskBlock for DNN ranking systems in this work, as shown in Figure \ref{Fig.figure-2a} and Figure \ref{Fig.figure-2b}. The proposed MaskBlock  consists of three key components: layer normalization module ,instance-guided mask, and a feed-forward hidden layer. The layer normalization can ease optimization of the network. The instance-guided mask introduces multiplicative interactions for feed-forward layer of a standard DNN model and feed-forward hidden layer aggregate the masked information in order to better capture the important feature interactions. In this way, we turn the widely used feed-forward layer of a standard DNN model into a mixture of addictive and multiplicative feature interactions.

First, we briefly review the formulation of LayerNorm.

\noindent\textbf{Layer Normalization:}\\
\noindent In general, normalization aims to ensure that signals have zero mean and unit variance as they propagate through a network to reduce "covariate shift" \cite{ioffe2015batch}. As an example, layer normalization (Layer Norm or LN)\cite{ba2016layer} was proposed to ease optimization of recurrent neural networks. Specifically, let $x = (x_1, x_2, ..., x_H)$ denotes the vector representation of an input of size $H$ to normalization layers. LayerNorm re-centers and re-scales input $\mathbf{x}$ as

\begin{equation}
  \setlength\abovedisplayskip{0pt}
  \begin{split}
  &\mathbf{h} = \mathbf{g} \odot N(\mathbf{x}) + \mathbf{b}, \quad N(\mathbf{x}) = \frac{\mathbf{x}-\mu}{\delta}, \\
  & \mu = \frac{1}{H}\sum^H_{i=1}x_i, \quad \delta = \sqrt{\frac{1}{H}\sum^{H}_{i=1}(x_i - \mu)^2}
\end{split}
\end{equation}
where $h$ is the output of a LayerNorm layer. $\odot$ is an element-wise production operation. $\mu$ and $\delta$ are the mean and standard deviation of input. Bias $\mathbf{b}$ and gain $\mathbf{g}$ are parameters with the same dimension $H$.

As one of the key component in MaskBlock, layer normalization can be used on both feature embedding and feed- forward layer. For the feature embedding layer, we regard each feature's embedding as a layer to compute the mean, standard deviation, bias and gain of LN as follows:
\begin{equation}
  LN\_EMB(\mathbf{V}_{emb}) = concate\left(LN(\mathbf{e}_1), LN(\mathbf{e}_2), ..., LN(\mathbf{e}_i), ..., LN(\mathbf{e}_f)\right)
\end{equation}

As for the feed-forward layer in DNN model, the statistics of $LN$ are estimated among neurons contained in the corresponding hidden layer as follows:

\begin{equation}
  LN\_HID(\mathbf{V}_{hidden}) = ReLU(LN(\mathbf{W}_i\mathbf{X}))
\end{equation}
where $\mathbf{X} \in \mathbb{R}^t$ refers to the input of feed-forward layer, $\mathbf{W}_i \in \mathbb{R}^{m \times t}$ are parameters for the layer, $t$ and $m$ respectively denotes the  size of input layer and neural number of feed-forward layer. Notice that we have two places to put normalization operation on the MLP: one place is before non-linear operation and another place is after non-linear operation. We find the performance of the normalization before non-linear consistently outperforms that of the normalization after non-linear operation. So all the normalization used in MLP part is put before non-linear operation in our paper as formula (4) shows.

\noindent\textbf{MaskBlock on Feature Embedding:}\\
\noindent We propose MaskBlock by combining the three key elements: layer normalization, instance-guided mask and a following feed-forward layer. MaskBlock can be stacked to form deeper network. According to the different input of each MaskBlock, we have two kinds of  MaskBlocks: MaskBlock on feature embedding and MaskBlock on Maskblock. We will firstly introduce the MaskBlock on feature embedding as depicted in Figure \ref{Fig.figure-2a} in this subsection.

The feature embedding $\mathbf{V}_{emb}$  is the only input for MaskBlock on feature embedding. After the layer normalization operation on embedding $\mathbf{V}_{emb}$. MaskBlock utilizes instance-guided mask to highlight the informative elements in $\mathbf{V}_{emb}$ by element-wise product, Formally,
\begin{equation}
  \begin{split}
  \mathbf{V}_{maskedEMB} = \mathbf{V}_{mask} \odot LN\_EMB(\mathbf{V}_{emb})
\end{split}
\end{equation}
 where $\odot$ means an element-wise production between the instance-guided mask and the normalized vector $LN_EMB(\mathbf{V}_{emb})$, $\mathbf{V}_{maskedEMB}$ denote the masked feature embedding. Notice that the input of instance-guided mask $\mathbf{V}_{mask}$ is also the feature embedding $V_{emb}$.

 \begin{figure*}
   \setlength{\abovecaptionskip}{1pt}
   \includegraphics[width=0.95\linewidth]{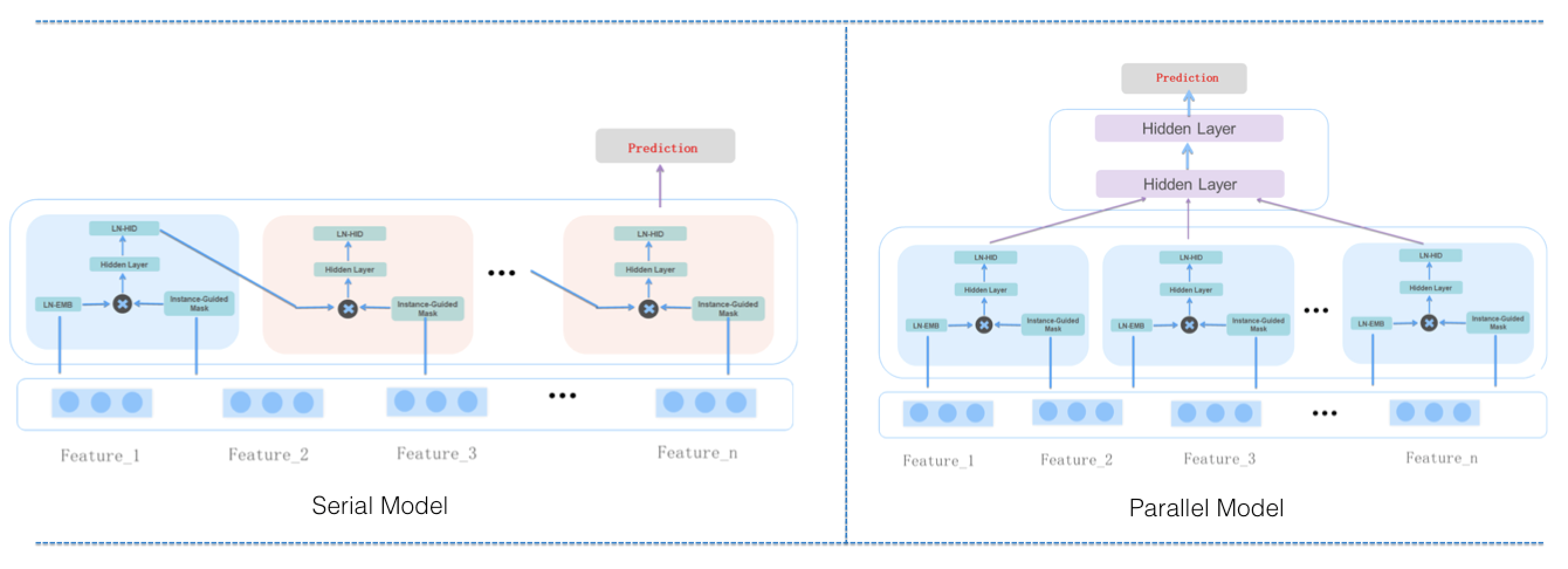}
   \caption{Structure of Serial Model and Parallel Model}
   \label{Fig.Structure}
 \end{figure*}

We introduce  a feed-forward hidden layer and a following layer normalization operation in MaskBlock to better aggregate the masked information by a normalized non-linear transformation. The output of MaskBlock can be calculated as follows:
\begin{equation}
  \begin{split}
\mathbf{V}_{output} &= LN\_HID(W_iV_{maskdEMB}) \\
&= ReLU(LN(\mathbf{W}_i(\mathbf{V}_{mask} \odot LN\_EMB(\mathbf{V}_{emb}))))
\end{split}
\end{equation}
where $\mathbf{W}_i \in \mathbb{R}^{q \times n}$ are parameters of the feed-forward layer in the $i$-th MaskBlock, $n$ denotes the size of $\mathbf{V}_{maskedEMB}$ and $q$ means the size of neural number of the feed-forward layer.

 The instance-guided mask introduces the element-wise product  into feature embedding  as a fine-grained  attention while normalization both on feature embedding and hidden layer eases the network optimization. These key components  in MaskBlock help the feed-forward layer capture complex feature cross more efficiently.

\noindent\textbf{MaskBlock on MaskBlock:}\\
In this subsection, we will introduce MaskBlock on MaskBlock as depicted in Figure \ref{Fig.figure-2b}. There are two different inputs for this MaskBlock: feature embedding $V_{emb}$ and the output $V_{output}^{p}$ of the previous MaskBlock. The input of instance-guided mask for this kind of MaskBlock is always the feature embedding $V_{emb}$. MaskBlock utilizes instance-guided mask to highlight the important feature interactions in previous MaskBlock's output $V_{output}^p$ by element-wise product, Formally,
\begin{equation}
  V_{maskedHID} = V_{mask} \odot V_{output}^p
\end{equation}
 where $\odot$ means an element-wise production between the instance-guided mask  $V_{mask}$ and the previous MaskBlock's output $V_{output}^p$, $V_{maskedHID}$ denote the masked hidden layer.

In order to better capture the important feature interactions, another feed-forward hidden layer and a following layer normalization are introduced in MaskBlock . In this way, we turn the widely used feed-forward layer of a standard DNN model into a mixture of addictive and multiplicative feature interactions to avoid the ineffectiveness of those addictive feature cross models.  The output of MaskBlock can be calculated as follows:
\begin{equation}
  \begin{split}
\mathbf{V}_{output} &= LN\_HID(W_iV_{maskdHID}) \\
&= ReLU(LN(\mathbf{W}_i(\mathbf{V}_{mask} \odot \mathbf{V}_{output}^p)))
\end{split}
\end{equation}
where $W_i \in \mathbb{R}^{q\times n}$ are parameters of the feed-forward layer in the $i$-th MaskBlock, $n$ denotes the size of $\mathbf{V}_{maskedHID}$ and $q$ means the size of neural number of the feed-forward layer.

\subsection{MaskNet}

Based on the MaskBlock, various new ranking models can be designed according to different configurations.  The rank model consisting of MaskBlock is called MaskNet in this work. We also propose two MaskNet models by utilizing the MaskBlock as the basic building block.

\noindent\textbf{Serial MaskNet: }\\
\noindent We can stack one MaskBlock after another to build the ranking system , as shown by the left model in  Figure \ref{Fig.Structure}. The first block is a MaskBlock on feature embedding and all other blocks are MaskBlock on Maskblock to form a deeper network. The prediction layer is put on the final MaskBlock's output vector. We call MaskNet under this serial configuration as SerMaskNet in our paper. All  inputs of instance-guided mask in every MaskBlock come from the feature embedding layer $\mathbf{V}_{emb}$ and this makes the serial MaskNet model look like a RNN model with sharing input at each time step.

\noindent\textbf{Parallel MaskNet: }\\
\noindent We propose another MaskNet by placing  several MaskBlocks on feature embedding in parallel on a sharing feature embedding layer, as depicted by the right model in Figure \ref{Fig.Structure}. The input of each block is only the shared feature embedding $\mathbf{V}_{emb}$ under this configuration. We can regard this ranking model as a mixture of multiple experts  just as  MMoE\cite{ma2018modeling} does. Each MaskBlock pays attention to specific kind of important features or feature interactions. We collect the information of each expert by concatenating the output of each MaskBlock as follows:
\begin{equation}
  \mathbf{V}_{merge} = concate(\mathbf{V}^1_{output}, \mathbf{V}^2_{output},...,\mathbf{V}^i_{output},...,\mathbf{V}^u_{output})
\end{equation}
where $\mathbf{V}^i_{output} \in \mathbb{R}^q$ is the output of the $i$-th MaskBlock and $q$ means size of neural number of feed-forward layer in MaskBlock, $u$ is the MaskBlock number.

To further merge the feature interactions captured by each expert, multiple feed-forward layers are stacked on the concatenation information $\mathbf{V}_{merge}$. Let $\mathbf{H}_0 = \mathbf{V}_{merge}$ denotes the output of the concatenation layer, then $\mathbf{H}_0$ is fed into the deep neural network and the feed forward process is:
\begin{equation}
  \mathbf{H}_l = ReLU(\mathbf{W}_l\mathbf{H}_{l-1} + \beta_l)
\end{equation}
where $l$ is the depth and ReLU is the activation function. $\mathbf{W}_t, \beta_t, \mathbf{H}_l$ are the model weight, bias and output of the $l$-th layer. The prediction layer is put on the last layer of multiple feed-forward networks. We call this version MaskNet as "ParaMaskNet" in the following part of this paper.

\subsection{Prediction Layer}
To summarize, we give the overall formulation of our proposed model’ s output as:
\begin{equation}
  \hat{y} = \delta(w_0 + \sum^n_{i=1}w_ix_i)
\end{equation}
where $\hat{y} \in (0, 1)$ is the predicted value of CTR, $\delta$ is the sigmoid function, $n$ is the size of the last MaskBlock's output(SerMaskNet) or feed-forward layer(ParaMaskNet), $x_i$ is the bit value of feed-forward layer and $w_i$ is the learned weight for each bit value.

For binary classifications, the loss function is the log loss:

\begin{equation}
  \mathcal{L} = -\frac{1}{N}\sum^N_{i=1}y_i\log(\hat{y}_i)+(1-y_i)\log(1-\hat{y}_i)
\end{equation}
where $N$ is the total number of training instances, $y_i$ is the ground truth of $i$-th instance and $\hat{y}_i$ is the predicted CTR. The optimization process is to minimize the following objective function:

\begin{equation}
\mathfrak{L} = \mathcal{L} + \lambda \|\Theta\|
\end{equation}
where $\lambda$ denotes the regularization term and $\Theta$ denotes the set of parameters, including those in feature embedding matrix, instance-guided mask matrix, feed-forward layer in MaskBlock, and prediction part.

\section{Experimental Results}
In this section, we evaluate the proposed approaches on three real-world datasets and conduct detailed ablation studies to answer the following research questions:

\begin{itemize}
\item\noindent\textbf{RQ1} Does the proposed MaskNet model based on the  MaskBlock perform better than existing state-of-the-art  deep learning based CTR models?
\item\noindent\textbf{RQ2} What are the influences of various components in the MaskBlock architecture? Is each component  necessary to build an effective ranking system?
\item\noindent\textbf{RQ3} How does the hyper-parameter of networks influence the performance of our proposed  two MaskNet models?
\item\noindent\textbf{RQ4} Does instance-guided mask highlight the important elements in feature embedding and feed-forward layers according to the input instance?
\end{itemize}

In the following, we will  first describe the experimental settings, followed by answering the above research questions.

\subsection{Experiment Setup}

\subsubsection{Datasets}

The following three data sets are used in our experiments:

\begin{enumerate}
  \item \textbf{Criteo\footnote{Criteo \url{http://labs.criteo.com/downloads/download-terabyte-click-logs/}} Dataset:}
  As a very famous public real world display ad dataset with each ad display information and corresponding user click feedback, Criteo data set is widely used in many CTR model evaluation. There are $26$ anonymous categorical fields and $13$ continuous feature fields in Criteo data set.

  \item \textbf{Malware\footnote{Malware \url{https://www.kaggle.com/c/microsoft-malware-prediction}} Dataset:}
  Malware is a dataset from Kaggle competitions published in the Microsoft Malware prediction. The goal of this competition is to predict a Windows machine's probability of getting infected. The malware prediction task can be formulated as a binary classification problem like a typical CTR estimation task does.

  \item \textbf{Avazu\footnote{Avazu \url{http://www.kaggle.com/c/avazu-ctr-prediction}} Dataset:}
    The Avazu dataset consists of several days of ad click- through data which is ordered chronologically. For each click data, there are $23$ fields which indicate elements of a single ad impression.
\end{enumerate}

We randomly split instances by $8:1:1$ for training , validation and test while Table \ref{tab:datasets} lists the statistics of the evaluation datasets.

\begin{table}[h]
  \setlength{\abovecaptionskip}{1pt}
\centering
\caption{Statistics of the evaluation datasets}
\begin{tabular}{lccc}
\toprule
Datasets  & \#Instances & \#fields & \#features \\
\midrule
Criteo       & 45M  & 39 & 30M     \\
Avazu       & 40.43M  & 23 & 9.5M     \\
Malware     & 8.92M  & 82 & 0.97M \\
\bottomrule
\end{tabular}
\label{tab:datasets}
\end{table}

\subsubsection{Evaluation Metrics}
AUC (Area Under ROC) is used in our experiments as the evaluation metric. AUC's upper bound is $1$ and larger value indicates a better performance.

RelaImp is also as work \cite{inproceedings} does to measure the relative AUC improvements over the corresponding baseline model as another evaluation metric. Since AUC is $0.5$ from a random strategy, we can remove the constant part of the AUC score and formalize the RelaImp as:
\begin{equation}
  RelaImp = \frac{AUC(Measured\ Model) - 0.5}{AUC(Base\ Model) - 0.5} - 1
\end{equation}

\begin{table*}
  \setlength{\abovecaptionskip}{1pt}
\centering
\caption{Overall performance (AUC) of different models on three datasets(feature embedding size=10,our proposed two models  both have 3 MaskBlocks  with same default settings.)}
\begin{tabular}{l|cccccc}
\toprule
  & \multicolumn{2}{c}{\textbf{Criteo}} &
  \multicolumn{2}{c}{\textbf{Malware}} & \multicolumn{2}{c}{\textbf{Avazu}} \\
\midrule
  & AUC & RelaImp   & AUC & RelaImp  & AUC & RelaImp \\
\midrule
FM & 0.7895 & 0.00\% & 0.7166 & 0.00\% & 0.7785 & 0.00\%\\
DNN & 0.8054 & +5.35\% & 0.7246 & +3.70\% & 0.7820 & +1.26\%\\
DeepFM & 0.8057 & +5.46\% & 0.7293 & +5.86\% & 0.7833 & +1.72\%\\
\midrule

DCN & 0.8058 & +5.49\% & 0.7300 & +6.19\% & 0.7830 & +1.62\%\\
xDeepFM & 0.8064 & +5.70\% & 0.7310 & +6.65\% & 0.7841 & +2.01\%\\
AutoInt & 0.8051 & +5.39\% & 0.7282 & +5.36\% & 0.7824 & +1.40\%\\
\midrule
SerMaskNet & 0.8119 & +7.74\% & \textbf{0.7413} & \textbf{+11.40\%} & \textbf{0.7877} & \textbf{+3.30\%}\\
ParaMaskNet & \textbf{0.8124} & \textbf{+7.91\%} & 0.7410 & +11.27\% & 0.7872 & +3.12\%\\
\bottomrule
\end{tabular}
\label{tab:overalperformance}
\end{table*}

\subsubsection{Models for Comparisons}
We compare the performance of the following CTR estimation models with our proposed approaches: FM, DNN, DeepFM, Deep\&Cross Network(DCN), xDeepFM and AutoInt Model, all of which are discussed in Section 2. FM is considered as the base model in evaluation.

\subsubsection{Implementation Details}

We implement all the models with Tensorflow in our experiments. For optimization method, we use the Adam with a mini-batch size of $1024$ and a learning rate is set to $0.0001$.  Focusing on neural networks structures in our paper, we make the dimension of field embedding for all models to be a fixed value of $10$. For models with DNN part, the depth of hidden layers is set to $3$, the number of neurons per layer is $400$, all activation function is ReLU. For default settings in MaskBlock, the reduction ratio of  instance-guided mask is set to $2$. We conduct our experiments with $2$ Tesla $K40$ GPUs.

\subsection{Performance Comparison (RQ1)}
The overall performances of different models on three evaluation datasets are show in the Table \ref{tab:overalperformance}. From the experimental results, we can see that:
\begin{enumerate}
  \item Both the serial model and parallel model achieve  better performance on all three datasets and obtains significant improvements over the state-of-the-art methods. It can boost the accuracy over the baseline FM by $3.12\%$ to $11.40\%$, baseline DeepFM by $1.55\%$ to $5.23\%$, as well as  xDeepFM baseline by $1.27\%$ to $4.46\%$. We also conduct a significance test to verify that our proposed models outperforms baselines with the significance level $\alpha = 0.01$.

  Though maskNet model lacks similar module such as CIN in xDeepFM to explicitly capture high-order feature interaction, it still achieves better performance because of the existence of MaskBlock. The experiment results imply that MaskBlock indeed  enhance DNN Model's ability of capturing complex feature interactions through introducing multiplicative operation into DNN models by instance-guided mask on the  normalized feature embedding and feed-forward layer.

  \item As for the comparison of the serial model and parallel model, the experimental results show comparable performance on three evaluation datasets. It explicitly proves that MaskBlock is an effective basic building unit for composing various high performance ranking systems.
\end{enumerate}

\subsection{Ablation Study of MaskBlock (RQ2)}
In order to better understand the impact of each component in MaskBlock, we perform ablation experiments over key components of  MaskBlock by only removing one of them  to observe the performance change, including mask module, layer normalization(LN) and feed-forward network(FFN). Table \ref{tab:table4} shows the results of our two full version MaskNet models and its variants removing only one component.

From the results in Table \ref{tab:table4}, we can see that removing either instance-guided mask or layer normalization will decrease  model's performance and this implies that both the instance-guided mask and layer normalization are necessary components in  MaskBlock for its effectiveness. As for the feed-forward layer in MaskBlock, its effect on serial model or parallel model shows difference. The Serial model’s performance dramatically degrades while it seems do no harm to parallel model if we remove the feed-forward layer in MaskBlock. We deem this implies that the feed-forward layer in MaskBlock is important for merging the feature interaction information after instance-guided mask. For parallel model, the multiple feed-forward layers above parallel MaskBlocks have similar function as feed-forward layer in MaskBlock does and this  may produce performance difference between two models when we remove this component.

\begin{table}[h]
  \setlength{\abovecaptionskip}{1pt}
  \caption{Overall performance (AUC) of MaskNet models removing different component in MaskBlock on Criteo dataset(feature embedding size=$10$, MaskNet model has $3$ MaskBlocks.)}
  \label{tab:table4}
  \begin{tabular}{lcc}
  \toprule
   Model Name & SerMaskNet & ParaMaskNet \\
   \midrule
   Full & 0.8119 & 0.8124 \\
   \hline
   -w/o Mask & 0.8090 & 0.8093 \\
   -w/o  LN & 0.8106 & 0.8103 \\
   -w/o FFN & 0.8085 & 0.8122 \\
   \bottomrule
\end{tabular}
\end{table}

\subsection{Hyper-Parameter Study(RQ3)}
In the following part of the paper, we study the impacts of hyper-parameters on two MaskNet models, including $1)$ the number of feature embedding size; $2)$ the number of MaskBlock; and $3)$ the reduction ratio in instance-guided mask module. The experiments are conducted on Criteo dataset via changing one hyper-parameter while holding the other settings. The hyper-parameter experiments show similar trend in other two datasets.

\paragraph{\textbf{Number of Feature Embedding Size.}}   The results in Table \ref{tab:table5} show the impact of the number of feature embedding size on model performance. It can be observed that the performances of  both models increase when embedding size increases at the beginning. However, model performance degrades when the embedding size is set greater than $50$ for SerMaskNet model and $30$ for ParaMaskNet model. The experimental results tell us the models benefit from larger feature embedding size.

\begin{table}
  \setlength{\abovecaptionskip}{1pt}
  \caption{Overall performance (AUC) of different feature embedding size of MaskNet Models on Criteo dataset(the number of MaskBlock is $3$)}
  \label{tab:table5}
  \begin{tabular}{lccccc}
  \toprule
  Embedding Size &    10 & 20 & 30 & 50 & 80 \\
  \midrule
  SerMaskNet & 0.8119 & 0.8123 & 0.8121 & 0.8125 & 0.8121 \\
  ParaMaskNet & 0.8124 & 0.8128 & 0.8131 & 0.8129 & 0.8129 \\
  \bottomrule
\end{tabular}
\end{table}

\paragraph{\textbf{Number of MaskBlock.}}   For understanding the influence of the number of MaskBlock on model's performance, we conduct experiments to stack MaskBlock from $1$ to $9$ blocks for both MaskNet models.  The experimental results are listed in the Table \ref{tab:table6}. For SerMaskNet model, the performance increases with more blocks at the beginning until the number is set greater than $5$. While the performance slowly increases when we continually add more MaskBlock into ParaMaskNet model. This may indicates that more experts boost the ParaMaskNet model's performance though it's more time consuming.

\begin{table}
  \setlength{\abovecaptionskip}{1pt}
  \caption{Overall performance (AUC) of different numbers of MaskBlocks in MaskNet model on Criteo dataset(embedding size$=10$)}
  \label{tab:table6}
  \begin{tabular}{lccccc}
  \toprule
  Block Number &    1 & 3 & 5 & 7 & 9 \\
  \midrule
  SerMaskNet & 0.8110 & 0.8119 & 0.8126 & 0.8117 & 0.8115 \\
  ParaMaskNet & 0.8113 & 0.8124 & 0.8127 & 0.8128 & 0.8132 \\
  \bottomrule
\end{tabular}
\end{table}

\begin{table}
  \setlength{\abovecaptionskip}{1pt}
  \caption{Overall performance (AUC) of different size of Hidden Layer in Mask Module of MastBlock on Criteo dataset.(embedding size$=10$, number of MaskBlock is $3$)}
  \label{tab:table7}
  \begin{tabular}{lccccc}
  \toprule
  Reduction   ratio &    1 & 2 & 3 & 4 & 5 \\
  \midrule
   SerMaskNet & 0.8118 & 0.8119 & 0.8120 & 0.8117 & 0.8119 \\
    ParaMaskNet & 0.8124 & 0.8124 & 0.8122 & 0.8122 & 0.8124 \\
  \bottomrule
\end{tabular}
\end{table}

\paragraph{\textbf{Reduction Ratio in Instance-Guided Mask.}} In order to explore the influence of the reduction ratio  in instance-guided mask, We conduct some experiments to adjust the reduction ratio from $1$ to $5$ by changing the size of aggregation layer.  Experimental results are shown in Table \ref{tab:table7} and we can observe that  various reduction ratio has little influence on model's performance. This indicates that we can adopt small reduction ratio in aggregation layer in real life applications for saving the computation resources.

\subsection{Instance-Guided Mask Study(RQ4)}
As discussed in Section in 3.2, instance-guided mask can be regarded as a special kind of bit-wise attention mechanism to highlight important information based on the current input instance. We can utilize instance-guided mask to boost the informative elements and suppress the uninformative elements or even noise in feature embedding and feed-forward layer.

To verify this, we design the following experiment: After training the SerMaskNet with $3$ blocks, we input different  instances into the model and observe the outputs of  corresponding instance-guided masks.

\begin{figure}
  \setlength{\abovecaptionskip}{1pt}
  \includegraphics[width=.7\linewidth]{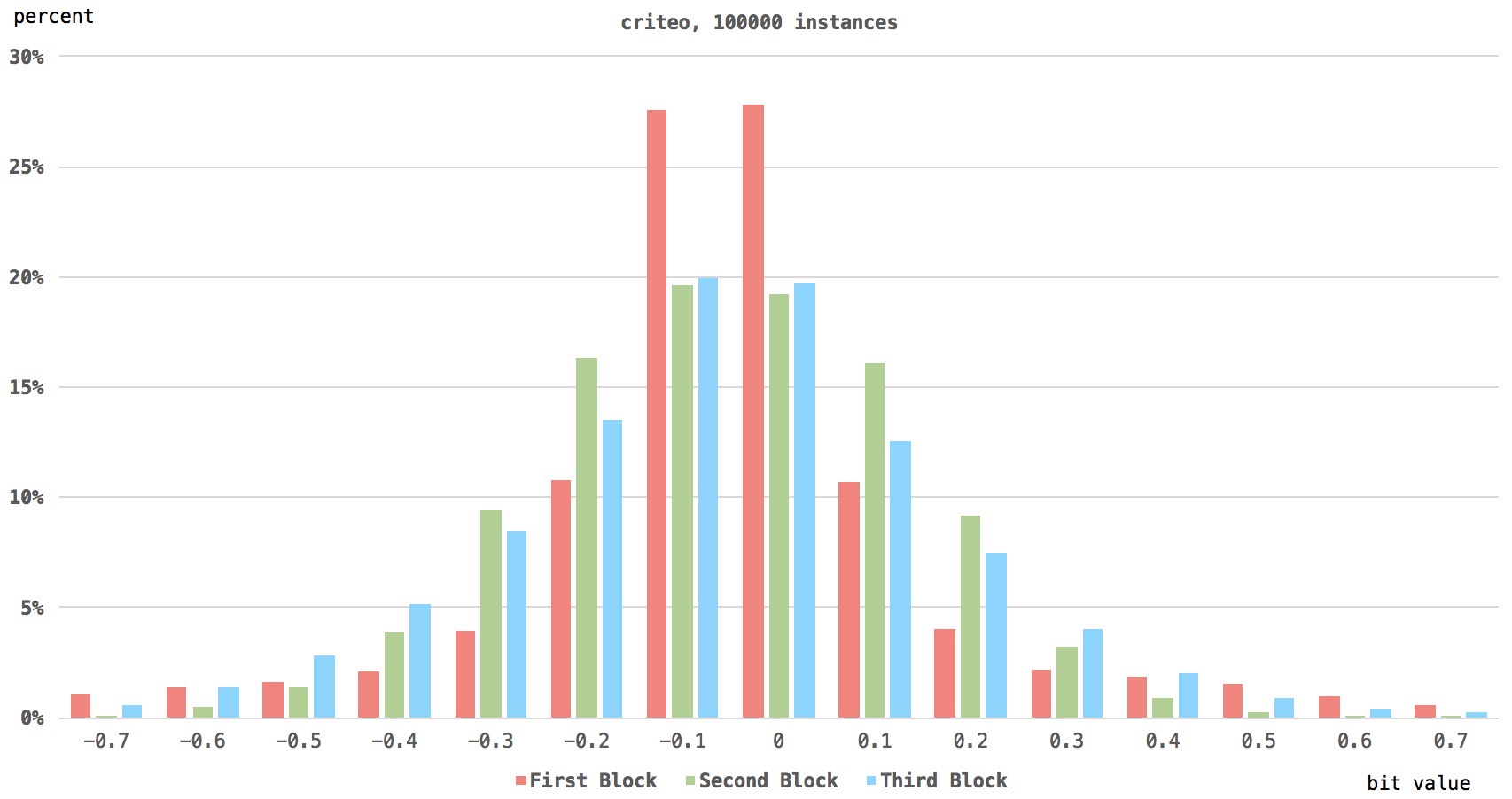}
  \caption{Distribution of Mask Values}
  \label{Fig.4}
\end{figure}

\begin{figure}
  \setlength{\abovecaptionskip}{1pt}
  \includegraphics[width=.49\linewidth]{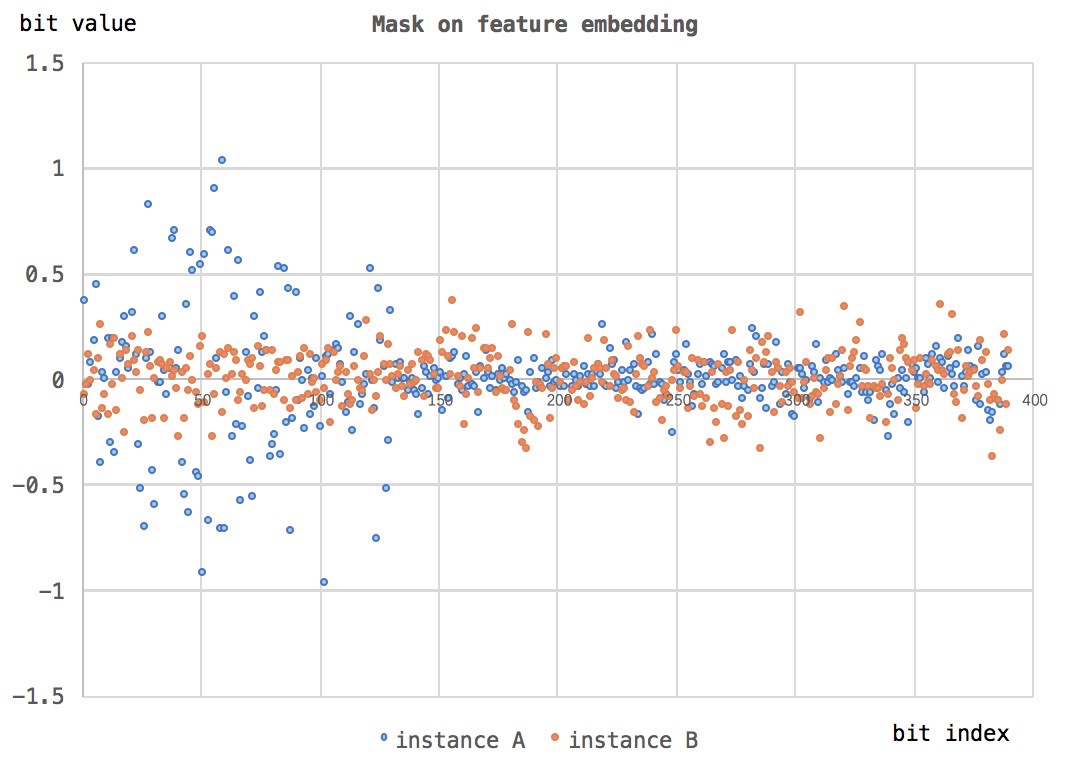}\hfill
  \includegraphics[width=.49\linewidth]{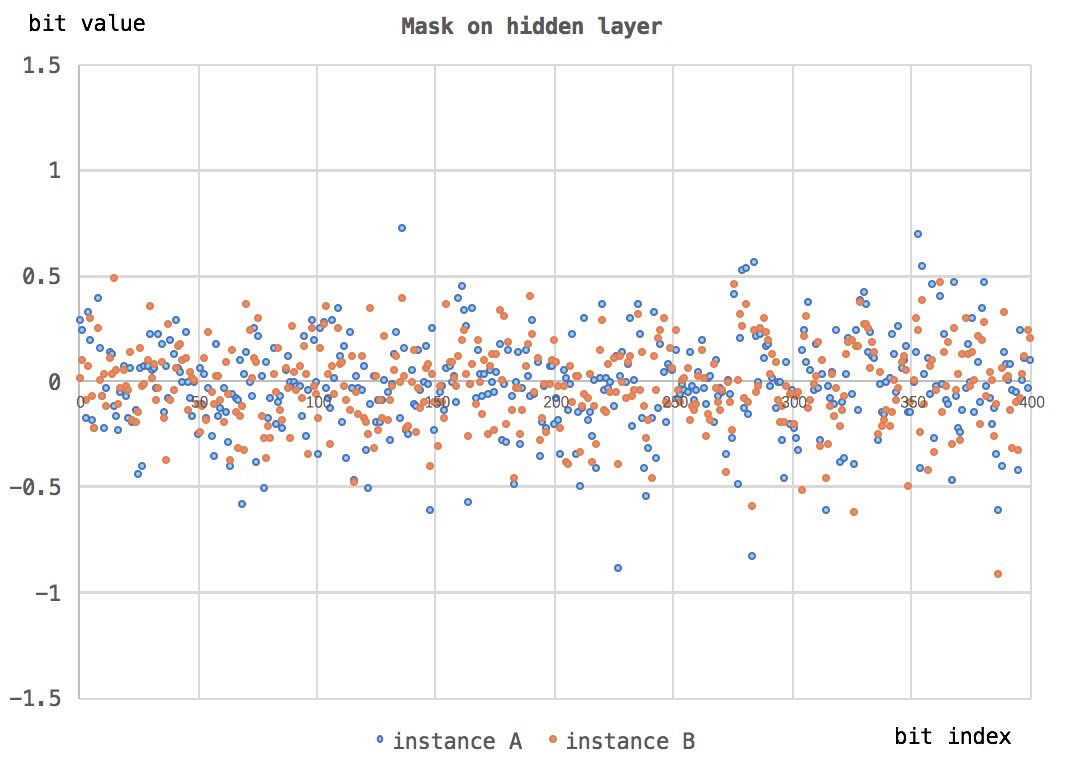}
  \caption{Mask Values of Two Expamples}
  \label{Fig.5}
\end{figure}

Firstly, we randomly sample $100000$ different instances from Criteo dataset and observe the distributions of the produced values by instance-guided mask from different blocks. Figure \ref{Fig.4} shows the result. We can see that the distribution of mask values follow normal distribution. Over 50\% of the mask values are small number near zero and only little fraction of the mask value is a relatively larger number. This implies that large fraction of  signals in feature embedding and feed-forward layer is uninformative or even noise which is suppressed by the small mask values. However, there is some informative information boosted by larger mask values through instance-guided mask.

Secondly, we randomly sample two instances and compare the difference of the produced values  by instance-guided mask. The results are shown in Figure \ref{Fig.5}. We can see that: As for the mask values for feature embedding, different input instances lead the mask to pay attention to various areas. The mask outputs of instance A pay more attention to the first few features and the mask values of instance B focus on some bits of other features. We can observe the similar trend in the mask values in feed-forward layer. This indicates the input instance indeed guide the mask to pay attention to the different part of the feature embedding and feed-forward layer.

\section{Conclusion}
In this paper, we introduce multiplicative operation into DNN ranking system by proposing  instance-guided mask which performs element-wise product both on the feature embedding and feed-forward layers. We also turn the feed-forward layer in DNN model into a mixture of addictive and multiplicative feature interactions by proposing MaskBlock by bombing the layer normalization, instance-guided mask, and feed-forward layer. MaskBlock is a basic building block to be used to design new ranking model. We also propose two specific MaskNet models based on the MaskBlock. The experiment results on three real-world datasets demonstrate that our proposed models outperform state-of-the-art models such as DeepFM and xDeepFM significantly.

\bibliographystyle{ACM-Reference-Format}
\bibliography{ref}
\end{document}